\documentclass[aps,pre,twocolumn,superscriptaddress,showpacs,tight]{revtex4-2}
\usepackage{amssymb}
\usepackage{epsfig}
\usepackage{amsmath}
\usepackage{times}
\usepackage{color}
\usepackage{lipsum}
\usepackage{ulem}
\usepackage{sidecap}
\begin{document}

\title{Criticality in Reservoir Computer of Coupled Phase Oscillators}

\author{Liang Wang}
\affiliation{School of Physics and Information Technology, Shaanxi Normal University, Xi'an 710062, China}
\author{Huawei Fan}
\affiliation{School of Physics and Information Technology, Shaanxi Normal University, Xi'an 710062, China}
\author{Jinghua Xiao}
\affiliation{School of Science, Beijing University of Posts and Telecommunications, Beijing 100876, China}
\author{Yueheng Lan}
\affiliation{School of Science, Beijing University of Posts and Telecommunications, Beijing 100876, China}
\author{Xingang Wang}
\email[Email address: ]{wangxg@snnu.edu.cn}
\affiliation{School of Physics and Information Technology, Shaanxi Normal University, Xi'an 710062, China}

\begin{abstract}
Accumulating evidences show that the cerebral cortex is operating near a critical state featured by power-law size distribution of neural avalanche activities, yet evidence of this critical state in artificial neural networks mimicking the cerebral cortex is lacking. Here we design an artificial neural network of coupled phase oscillators and, by the technique of reservoir computing in machine learning, train it for predicting chaos. It is found that when the machine is properly trained, oscillators in the reservoir are synchronized into clusters whose sizes follow a power-law distribution. This feature, however, is absent when the machine is poorly trained. Additionally, it is found that despite the synchronization degree of the original network, once properly trained, the reservoir network is always developed to the same critical state, exemplifying the ``attractor" nature of this state in machine learning. The generality of the results is verified in different reservoir models and by different target systems, and it is found that the scaling exponent of the distribution is independent on the reservoir details and the bifurcation parameter of the target system, but is modified when the dynamics of the target system is changed to a different type. The findings shed lights on the nature of machine learning, and are helpful to the design of high-performance machine in physical systems. 
\end{abstract}

\maketitle

Model-free prediction of chaotic systems by the technique of reservoir computer (RC) in machine learning has received considerable attention in recent years~\cite{RC:Jaeger,RC:Lu2017,RC:Pathak2017,RC:Pathak2018,RC:Fan,RC:CK2020,RC:Tang2020,RC:FHW2021,RC:Patel2021}. From the perspective of dynamical systems, RC can be regarded as a complex network of coupled nonlinear units which, driven by the input signals, generate the output through a readout function~\cite{RC:lukosevicius2009}. In the training phase, the input signals are provided by the target system, and the purpose of the training is to find the set of coefficients in the readout function for a best fitting of the training data. In the predicting phase, the input signals are replaced by the outputs, and the machine is running as an autonomous system with fixed parameters. Although structurally simple, RC has shown its super power in many data-oriented applications~\cite{RC:Jaeger,RC:Lu2017,RC:Pathak2017,RC:Pathak2018,RC:Fan,RC:CK2020,RC:Tang2020,RC:FHW2021,RC:Patel2021,RC:lukosevicius2009,RC:Tanaka2019}, e.g., speech recognization, channel equalization, robot control and chaos prediction. In particular, it has been shown that a properly trained RC is able to predict accurately the state evolution of a chaotic system for about half a dozen Lyapunov times~\cite{RC:Jaeger,RC:Lu2017,RC:Pathak2017,RC:Pathak2018}, which is much longer than the prediction horizon of the traditional methods in nonlinear science. Besides predicting short-term state evolution, RC is also able to replicate faithfully the long-term statistical properties of chaotic systems, e.g., the attractors and Lyapunov exponents~\cite{RC:Pathak2017}. This ability, known as climate replication, has been exploited recently to reconstruct the bifurcation diagram of chaotic systems~\cite{KLW:2021,RC:FHW2021}, where it is shown that by introducing a parameter-control channel, a RC trained by the time series acquired at several bifurcation parameters is able to reproduce the entire bifurcation diagram. 

Whereas the capability of RC in predicting chaos has been well demonstrated, the principles for designing RC and the working mechanism of the machine remain elusive~\cite{RC:LZX2018,RC:Carroll2019,RC:AHChaos2019,RC:LYC2019,RC:Griffith2019,RC:LZX2020,RC:Carroll2020-2,RC:Carroll2020-3,RC:Herteux2020,RC:Bollt2021,LCH:EdgeChaos}. In designing RC, the general requirement is that the dynamics of the reservoir network should be complex enough~\cite{RC:Maass2002}, yet the performance of RC does depend on a number of factors, e.g., the network connectivity~\cite{RC:Carroll2019,RC:AHChaos2019,RC:Griffith2019}, the network spectral radius~\cite{RC:LYC2019}, the mean path length~\cite{RC:Carroll2020-2}, the symmetry properties~\cite{RC:Herteux2020}, etc. To give the optimal performance, all these factors should be taken into account and the corresponding parameters should be chosen very carefully, as a small change in any of the parameters might lead to a significant decrease in the prediction performance. This makes the design of high-performance RC a somewhat tricky task relying on experience and luck. An alternative approach to improve the performance of RC is exploring the working mechanism of the machine~\cite{RC:LZX2018,RC:LZX2020,RC:Carroll2020-3,RC:Bollt2021,LCH:EdgeChaos}, with the belief that the function of RC is supported by some generic dynamical features and a proper characterization of these features could provide useful guidance for setting the hyperparameters. This approach, however, is proven to be also uneven. One reason is that the collective behavior of the reservoir is dependent on both the target system and the specific machine function. Another reason is that the dynamics of the reservoir network shows a diversity of properties, which can hardly be unified to and explained by a single mechanism. Although not being able to deliver direct instructions for optimizing the hyperparameters, these studies do give useful suggestions on how to design a high-performance RC and, more importantly, shed lights on the nature of machine learning. For instance, it has been demonstrated that the computational capacity of RC is maximized at ``the edge of chaos (stability)"~\cite{RC:NB2004,RC:Legenstein2007,LCH:EdgeChaos}, namely the critical state at the transition from the ordered to chaotic states, and, to achieve the optimal performance, the machine should be tuned to the vicinity of this critical state. This finding is consistent with the findings in cellular automata~\cite{RC:Langton1990,SOC:Book}, and is widely regarded as one of the important principles for designing RC~\cite{RC:lukosevicius2009,RC:Choi2019,RC:Mandal2021,RC:Thomas2021}. This principle, however, has been questioned by some recent studies~\cite{RC:Carroll2020-3}, which show that for some types of nodal dynamics the prediction performance might be deteriorated when the reservoir is close to the critical state. 

As emulators of the human brain, artificial neural networks draw many inspirations from the brain in structure and dynamics~\cite{DeepLearning:Book}. Whereas the structural properties of the brain has been well exploited in designing artificial neural networks, the dynamical features of the brain have been largely overlooked. For instance, a signature of the brain dynamics is that the neurons are self-organized into a critical state characterized by a power-law size distribution of the neuronal events, namely the phenomenon of self-organized criticality~\cite{CritiExp:Beggs2003,Criti:Arcangelis2006,CritiExp:Pasquale2008,Criti:Friedman2012,Criti:Fontenele2019,Criti:ZCS2020}. This phenomenon, which is commonly regarded as the dynamical basis for many brain functions~\cite{Criti:Kinouchi2006,Criti:Beggs2008,Criti:Shew2009,CritiExp:FJF2020}, has not been reported in artificial neural networks. One possible reason is that, to improve the computing efficiency, nodes in artificial neural networks normally do not have intrinsic dynamics. Specifically, in the absence of input signals, all nodes are staying at their steady states. In contrast, neurons in the human brain are active and dynamic, in the sense that the state of an isolated neuron is able to oscillate with time without stimulations (the oscillatory neurons) or evolve for a transient period after being stimulated (the excitable neurons)~\cite{Neuroscience:Book}. This dynamical feature is crucial for neurons to be self-organized into various spatiotemporal patterns~\cite{Criti:Fontenele2019,Criti:ZCS2020,Criti:Santo2018,Criti:Porta2019}, in which a power-law scaling is found in the size distribution of the neural events. Regarding the significance of criticality to the brain functions, a natural question raised in machine learning is: Can criticality be observed in artificial neural networks and, if yes, what is the role it plays in machine learning? In the present work, by reservoir networks of coupled phase oscillators, we investigate the spatiotemporal patterns emerged in the reservoir by training the machine to predict the evolution of chaotic systems. Our main finding is that when the machine is properly trained (i.e., predicting accurately the system evolution for several Lyapunov times and replicating faithfully the dynamics climate), nodes in the reservoir are synchronized into clusters, with the sizes of the clusters following a power-law distribution -- a signature of criticality. In addition, we find that, once the machine is properly trained, the reservoir is always ``attracted" to the same critical state, despite the variations of the system parameters and the coherence degree of the original reservoir network.

The RC employed here is identical to the conventional ones in architecture~\cite{RC:Jaeger,RC:lukosevicius2009}, but with different reservoir dynamics. Specifically, the RC consists of three modules: the input layer, the reservoir network, and the output layer. The input layer is characterized the matrix $\mathbf{H}_{in}\in\mathbb{R}^{N\times D_{in}}$, which couples the input vector $\mathbf{u}(t)\in\mathbb{R}^{D_{in}}$ to the reservoir network. The elements of $\mathbf {H}_{in}$ are randomly drawn from the interval $[-1, 1]$. The reservoir network consists of $N$ non-identical phase oscillators, with the dynamics of the $i$th oscillator described by the equation
\begin{equation}
\begin{aligned}
\dot{\theta}_{i}(t)={(1-\alpha)\omega_{i}+\frac{\alpha}{N}\sum\limits_{j=1}^{N}a_{ij}\sin[\theta_{j}(t)-\theta_{i}(t)]} \\
+\beta\tanh\left[b_i+\sum\limits_{j=1}^{D_{in}}h_{ij}u_{j}(t)\right].\\
 \end{aligned}
 \label{reservoir-1}
\end{equation}    
Here $\theta_{i}(t)$, $\dot{\theta}_i(t)$ and $\omega_i$ denote, respectively, the instant phase, angular frequency and the natural frequency of the $i$th oscillator. The value of $\omega_i$ is randomly chosen within the range $(-1,1)$, and the initial phases of the oscillators are randomly chosen within the range $[-\pi,\pi)$. The coupling relationship of the oscillators is described by the matrix $\mathbf{A}\in\mathbb{R}^{N\times N}$, whose elements are randomly chosen from the interval $(0,1)$. $\alpha$ is the generalized coupling parameter used to adjust the coherence degree of the reservoir network. $\beta$ denotes the coupling strength of the input data. $\mathbf{b}\in\mathbb{R}^N$ is the bias vector of the input, whose elements are also randomly chosen from the interval $(0,1)$. The output layer is characterized by the matrix $\mathbf{H}_{out}\in\mathbb{R}^{D_{out}\times (N+D_{in}+1)}$, which generates the output vector $\mathbf{v}(t)\in \mathbb{R}^{D_{out}}$ by the operation
\begin{equation}
\mathbf{v}(t)=\mathbf{H}_{out}[\mathbf{b}_{out};\mathbf{u}(t);\mathbf{\Theta}(t)],
 \label{reservoir-2}
\end{equation}
with $\mathbf{\Theta}(t)\in \mathbb{R}^{N}$ the state vector of the reservoir network and $\mathbf{b}_{out}\in \mathbb{R}^N$ the bias vector of the outputs. For simplicity, the elements of $\mathbf{b}_{out}$ are set uniformly as $b_{out}=1$. Except the output matrix $\mathbf{H}_{out}$, all parameters in Eqs.~(\ref{reservoir-1}) and (\ref{reservoir-2}) are fixed at the construction. In simulations, Eq.~(\ref{reservoir-1}) is solved by the $4$th-order Runge-Kutta method, with the time step being set as $\delta t=5\times 10^{-2}$. 

The implementation of the RC contains three phases. First, the reservoir is evolving according to Eq.~(\ref{reservoir-1}) for a transient period $T_0$, so as to remove the influence of the initial conditions of the oscillators. Then, the reservoir is evolving for a period $T$ to estimate the output matrix $\mathbf{H}_{out}$, namely the training phase. The purpose of the training is to find a suitable matrix $\mathbf{H}_{out}$ so that the output vector $\mathbf{v}(t)$ as calculated by Eq.~(\ref{reservoir-2}) is as close as possible to the input vector $\mathbf{u}(t+\Delta t)$ for a time series of length $L$, with $\Delta t$ the time interval of data sampling. This can be done by minimizing a cost function with respect to $\mathbf{H}_{out}$, which gives
\begin{equation}
\mathbf{H}_{out}=\mathbf{U}\mathbf{V}^T(\mathbf{V}\mathbf{V}^T+\lambda \mathbb{I})^{-1},
\label{training}
\end{equation}
where $\mathbf{U}\in \mathbb{R}^{D_{out}\times L}$ is a matrix whose $k$th column is $\mathbf{u}[(k+1)\Delta t]$, $\mathbf{V}\in \mathbb{R}^{(D_{in}+N+1)\times L}$ is a matrix whose $k$th column is $[b_{out};\mathbf{u}(k\Delta t);\mathbf{\Theta}(k\Delta t)]^T$, $\mathbb{I}$ is the identity matrix, and $\lambda$ is the ridge regression parameter for avoiding the overfitting. In general, the input and output vectors are of different dimensions. For simplicity, here we set $D_{in}=D_{out}$. Finally, in the predicting phase, the input vector $\mathbf{u}(t)$ in Eq.~(\ref{reservoir-1}) is replaced by the output vector $\mathbf{v}(t+\delta t)$, and the machine is evolving as an autonomous system with the fixed matrix $\mathbf{H}_{out}$, with $\mathbf{v}(t)$ the predicted state at time $t$. 

\begin{figure*}[tbp]
\centering
\includegraphics[width=0.9\linewidth]{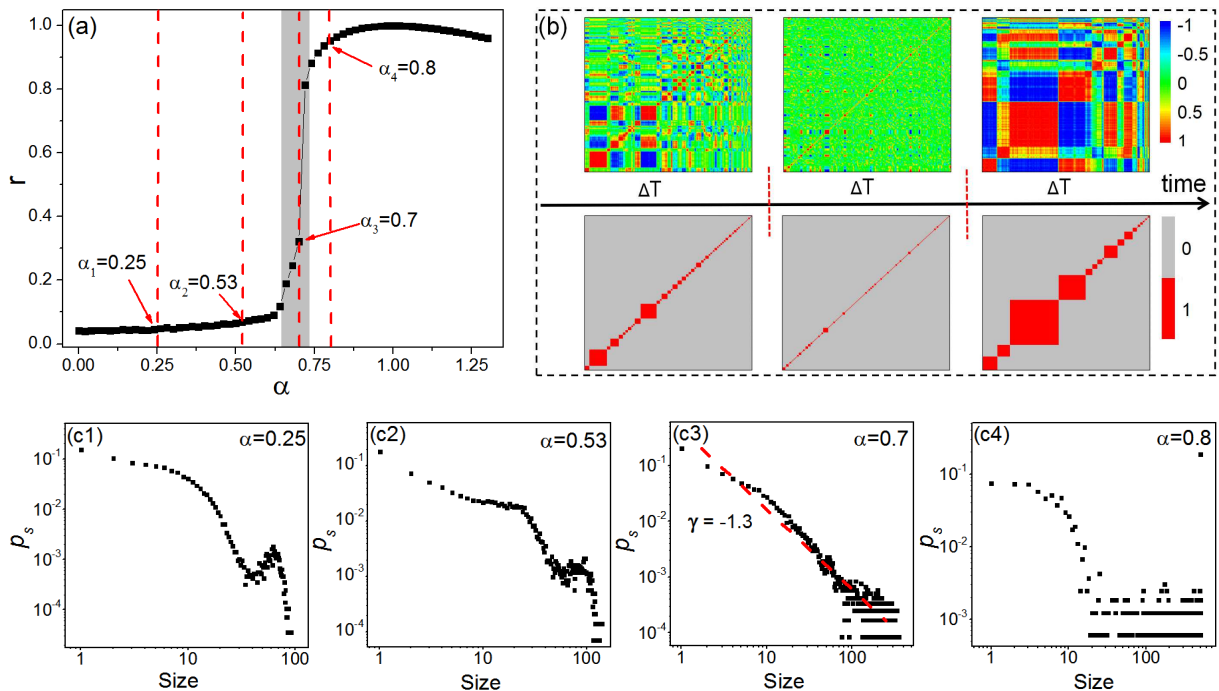}
\caption{The collective behavior of the reservoir network in the absence of input. (a) The variation of the order parameter $r$ with respect to the generic coupling strength $\alpha$. The grey zone, $\alpha\in(0.65,0.71)$, denotes the transition regime in which the value of $r$ is quickly increased with $\alpha$. The values of $r$ at $\alpha=0.25$, $0.53$, $0.7$ and $0.8$ are about $0.046$, $0.071$, $0.322$ and $0.952$, respectively. (b) For $\alpha=0.53$, typical states observed in the system evolution. Upper panels: the correlation matrices. Lower panels: the synchronization patterns. (c) The size distribution of the synchronization clusters with different coupling strengths. The size distribution in (c3) follows roughly a power-law scaling $p_s\propto s^{\gamma}$, with the fitted exponent $\gamma\approx -1.3$.}
\label{fig1}
\end{figure*}

Before using the new RC to predict the evolution of chaotic systems, we first explore the collective behavior of the reservoir network in the absence of inputs. This is done by setting $\beta=0$ in Eq.~(\ref{reservoir-1}). In this way, the dynamics of the reservoir is governed by the generalized Kuramoto model~\cite{KuramotoModel}, with $\alpha$ the generic coupling strength. (The results for the conventional Kuramoto model will be reported later.) We characterize the collective behavior of the reservoir by two statistical quantities: the order parameter and the size distribution of temporal synchronization clusters. The order parameter is defined as $r=\left< \left | \sum_{j=1}^N e^{i\theta_j}/N\right | \right>$, with $j$ the oscillator index, and $|\cdot|$ and $\left<\cdot\right>$ denote, respectively, the module and time-average functions. The value of $r$ ranges between 0 and 1, with $r=0$ and $1$ corresponding to the completely desynchronized and synchronized states, respectively. Setting $N=500$, we plot in Fig.~\ref{fig1}(a) the variation of $r$ with respect to $\alpha$. We see that $\alpha$ is staying at small values for $\alpha<0.65$, but is increased quickly as $\alpha$ increases from $0.65$ to $0.71$, and reaches $1$ at $\alpha=1$. (As $\alpha$ increases from $1$ further, the frequency mismatch between the oscillators is enlarged and the value of $r$ is decreased.) According to the behavior of $r$, the transition process can be roughly divided into three regimes: the desynchronization regime ($0<\alpha<0.65$), the transition regime ($0.65<\alpha<0.71$), and the strong synchronization regime ($0.71<\alpha<1$). The order parameter characterizes only the overall coherence degree of the reservoir, yet providing no information on the spatial and temporal properties. For the intermediate states in synchronization transition, a common phenomenon is that the oscillators are synchronized into clusters of different sizes and, during the course of system evolution, the contents and sizes of the clusters are continuously changed~\cite{SynPattern:ZCS2006,SynPattern:Wang2010,SynPattern:Choudhary2017,SynPattern:Kaneko2016,SynPattern:Gerster2020}, i.e., the non-stationary synchronization patterns. The properties of the patterns not only determine the synchronization degree, but also unveil the organizing mechanism of the oscillators in the transition from the desynchronized to synchronized states. We next study the collective behavior of the reservoir from the perspective of synchronization patterns, and investigate the variation of the pattern properties in synchronization transition.  

The synchronization patterns are calculated from the time series of the oscillators as follows. First, we truncate the whole time series into segments of equal length $\Delta T$, and calculate for each segment the pair-wise correlation $p_{ij}=\sum_{k}[\theta_i(k) \theta_j(k)]/[\sum_k( \theta_i (k))^2( \theta_j (k))^2]^{1/2}$, with $k=1,\ldots,\Delta T$ the index of the data points and $\theta_i(k)$ the phase state of the $i$th oscillator in the reservoir. We have $p\in[-1,1]$, with $p=-1$ and $1$ correspond to, respectively, the anti-phase and in-phase synchronization states. As an example, we set $\alpha=0.53$ and $\Delta T=260$, and plot in Fig.~\ref{fig1}(b) (the upper row) three successive correlation matrices observed in the evolution. Clearly, the correlation matrix is evolving with time. Next, we transform each correlation matrix $\mathbf{P}$ to a binary matrix $\mathbf{B}$ by setting a threshold $p_c$: $b_{ij}=1$ if $p_{ij}\geq p_c$ and $b_{ij}=0$ if $p_{ij}< p_c$. Finally, we check for each row of the matrix $\mathbf{B}$ the non-zero elements, which defines a synchronization cluster. The size of the cluster containing oscillator $i$ therefore is $s_l=\sum_i b_{ij}$, and we have $\sum_l s_l=N$, with $l=1,\ldots,n$ the cluster index. Note that $s=1$ corresponds to the trivial case of isolated oscillator, i.e., the oscillator is not synchronized with any other oscillator in the network. Setting $p_c=0.95$, we plot in Fig.~\ref{fig1}(b) (the lower row) the binary matrices transformed from the correlation matrices. We see that both the contents and sizes of the clusters are changing with time. 

\begin{figure*}[tbp]
\includegraphics[width=0.92\linewidth]{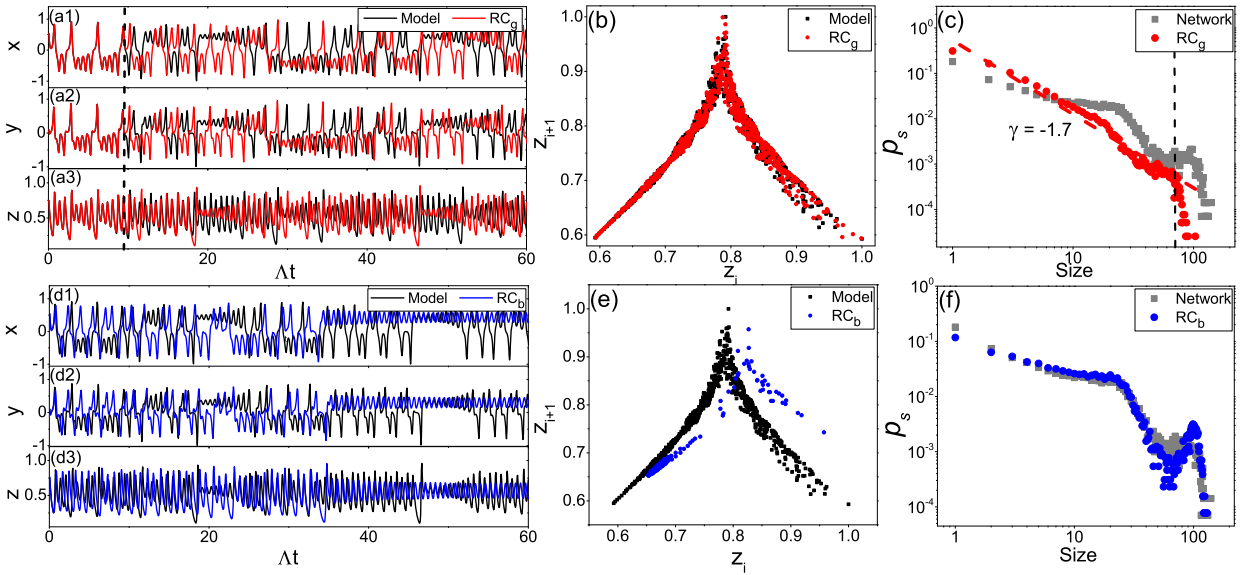}
\caption{By the coupling strength $\alpha=0.53$, the prediction performance and pattern properties of the properly ($RC_g$) and poorly ($RC_b$) trained machines. The parameters of $RC_g$ are $(\beta,\lambda)=(0.477,1\times 10^{-7})$. The parameters of $RC_b$ are $(\beta,\lambda)=(0.498,2\times 10^{-5})$. (a) The state evolution of the Lorenz chaotic oscillator predicted by $RC_g$. The prediction horizon is about $10$ Lyapunov times. $\Lambda\approx 1.03$ is the largest Lyapunov exponent of the chaotic Lorenz oscillator. (b) The return map of the variable $z$ predicted by $RC_g$. (c) The size distribution of the synchronization clusters in the predicting phase for $RC_g$. Grey squares: the results for the original network without the input [see Fig.~\ref{fig1}(c2)]. Red discs: the results for the reservoir network of $RC_g$. The results within the range $s\in [1,70]$ are fitted by a power-law scaling $p_s\propto s^{\gamma}$, with $\gamma\approx -1.7$. (d) The state evolution of the chaotic Lorenz oscillator predicted by $RC_b$. (e) The return map of the variable $z$ predicted by $RC_b$. (f) The size distribution of the synchronization clusters in the predicting phase for $RC_b$.}
\label{fig2}
\end{figure*}

We next characterize the reservoir dynamics by the size distribution of the synchronization clusters. In simulations, this is done by counting the number of clusters of size $s$ in $500$ successive patterns (segments). The probability of finding a cluster of size $s$ is denoted as $p_s$. Figure~\ref{fig1}(c) shows the variation of $p_s$ with respect to $s$ for different values of $\alpha$. We see that the distribution is dependent on $\alpha$. In particular, the distribution for $\alpha=0.7$ follows roughly a power-law scaling $p_s\propto s^{\gamma}$, with the fitted exponent $\gamma\approx-1.3$. The power-law scaling reveals the scale-free nature of the clusters, and is a hallmark of criticality in neuronal and nonlinear sciences~\cite{Criti:Fontenele2019,Criti:ZCS2020,Criti:Santo2018,Criti:Porta2019,SynPattern:Kaneko2016}. This feature, however, is missing for other values of $\alpha$ ($0.25$, $0.53$ and $0.8$) in Fig.~\ref{fig1}(c). [The feature of criticality is commonly observed in the transition regime $\alpha\in (0.65,0.71)$. See Supplementary Material for details.] In what follows, we are going to demonstrate that, despite the value of $\alpha$, once the machine is properly trained and is able to predict the evolution of chaotic system, the size distribution of the clusters will be always ``shaped" to the same power-law scaling, which, according to the studies in neuroscience, maximizes the computational capability and information transmission efficiency of the brain~\cite{Criti:Kinouchi2006,Criti:Beggs2008,Criti:Shew2009,CritiExp:FJF2020}.  

To demonstrate, we set $\alpha=0.53$ in the reservoir (within the desynchronization regime) and make it a RC by incorporating the input and output layers. The target system to be learned is the classical Lorenz oscillator: $(dx/dt,dy/dt,dz/dt)^{T}=[10(y-x),\rho x-y-xy,-8/3z+xy]^{T}$. We choose the parameter $\rho=35$, by which the system dynamics is chaotic. The time step in simulating the Lorenz oscillator is $\delta t=0.01$, and the system state $\mathbf{u}=(x,y,z)^T$ is acquired every $5$ time steps. As such, the time interval between successive data points is $\Delta t=0.05$, which is equal to the time step in updating the reservoir. (This setting is for convenience, but is not necessary.) The state variables are normalized to be within the range $[-1,1]$. Among the data, a segment of $T_0=1\times 10^3$ points is used to drive the reservoir out of the transient, and another segment of $T=2\times 10^3$ is used to estimate the output matrix $\mathbf{H}_{out}$ according to Eq.~(\ref{training}). The other parameters of the RC are $(\beta,\lambda)=(0.477,1\times 10^{-7})$, which are obtained by the  optimizer ``optimoptions" in MATLAB. Figure~\ref{fig2}(a) shows the state evolution of the oscillator predicted by the trained machine, together with the one obtained from model simulations. We see that the machine is able to predict accurately the system evolution for about $10$ Lyapunov times. Figure~\ref{fig2}(b) shows the return map of the variable $z$ predicted by the machine, which is well overlapped with the one obtained from model simulations, indicating that the climate of the oscillator has been successfully replicated~\cite{RC:Pathak2017}. In the present work, we regard a RC as being properly trained if it is able to predict the state evolution accurately (prediction error less than $5\%$) for several Lyapunov times, and denote it as $RC_{g}$. By tracing the evolutions of the synchronization patterns in the predicting phase, we plot in Fig.~\ref{fig2}(c) the size distribution of the synchronization clusters in the reservoir. We see that, different from the results of the original network [see Fig.~\ref{fig1}(c2)], the distribution now follows roughly a power-law scaling $p_s\propto s^{\gamma}$, with the fitted exponent $\gamma\approx -1.7$. (The normalized fitting error is about $0.28$. See Supplementary Material for details).  That is, by a proper training of the machine, the dynamics of the reservoir network is adjusted from a non-critical state to a critical state. The order parameter of the reservoir network now becomes $r=0.04$, which is smaller than that of the original network ($r=0.071$). The decreased order parameter is attributed to the breaking of the giant cluster in the reservoir. Specifically, a giant cluster containing a large portion of the oscillators is formed in the original network, which is broken into many medium-size clusters when the machine is properly trained. (See Supplementary Material for more details).  

How about the poorly trained machine? To show an example, we keep the other parameters of the RC unchanged but setting $(\beta,\lambda)=(0.498,2\times 10^{-5})$, and train the machine by the same set of input data. The state evolution predicted by the machine is plotted in Fig.~\ref{fig2}(d). We see that the machine is able to predict the evolution for only about $1$ Lyapunov time. Figure~\ref{fig2}(e) shows the return map constructed from the variable $z$ predicted by the machine, which is clearly different from the one obtained from model simulations. According to our definition of RC, we regard this machine as being poorly trained, and denote it by $RC_b$. The size distribution of the clusters in the predicting phase for the poorly trained machine is plotted in Fig.~\ref{fig2}(f). We see that the distribution is similar to that of the original network [Fig.~\ref{fig1}(c2)], both are clearly diverged from the power-law distribution. 

It is worth mentioning that the above features are independent on the parameters $\Delta T$ (the time window over which pair-wise correlation is calculated) and $p_c$ (the correlation threshold defining the synchronization clusters). In calculating the correlation matrices in Figs.~\ref{fig1} and \ref{fig2}, the value of $\Delta T$ is set as $260$, which corresponds to about $12$ Lyapunov times and is comparable to the prediction horizon of $RC_g$. Additional simulation shows that, given $\Delta T$ is not very small or large (e.g., varying $\Delta T$ within the range $[200,300]$), the size distribution of the clusters follows the same power-law scaling. The similar results are also observed when $p_c$ is changed. However, for the poorly trained machine $RC_b$, despite the variations of $\Delta T$ and $p_c$, the size distribution is always diverged from the power-law scaling. (See Supplementary Material for details.)   

\begin{figure}[tbp]
\includegraphics[width=0.9\linewidth]{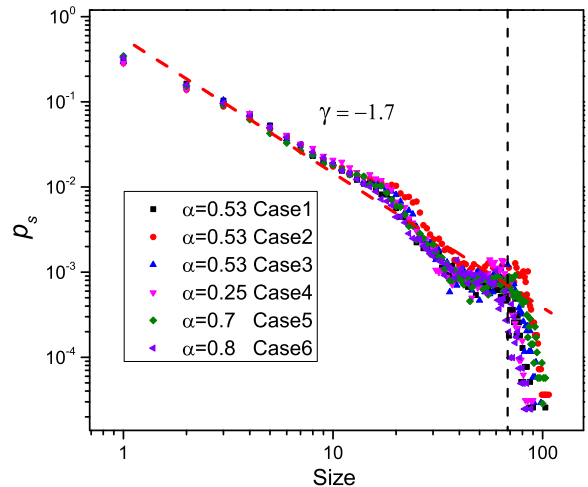}
\caption{The scaling property of the synchronization clusters in properly trained machine with different parameters $\alpha$, $\beta$ and $\lambda$. Case 1: $\alpha=0.53$, $(\beta,\lambda)=(0.477,1\times 10^{-7})$ [the reference distribution copied from Fig.~\ref{fig2}(c)]. Case 2: $\alpha=0.53$, $(\beta,\lambda)=(0.254,3\times 10^{-7})$. Case 3: $\alpha=0.53$, $(\beta,\lambda)=(0.371,1\times 10^{-7})$. Case 4: $\alpha=0.25$, $(\beta,\lambda)=(0.215,1\times 10^{-7})$. Cast 5: $\alpha=0.7$, $(\beta,\lambda)=(0.448,1\times 10^{-7})$. Case 6: $\alpha=0.8$, $(\beta,\lambda)=(0.616,1\times 10^{-7})$. For $s\in [1,70]$, all distributions follows the same power-law scaling $p_s\propto s^{\gamma}$, with $\gamma\approx -1.7$.}
\label{fig3}
\end{figure}

How about other parameters, saying, for example, changing the values of $\alpha$ (the generic coupling strength), $\beta$ (the input coupling strength) and $\lambda$ (the regression parameter)? We check first the impact of $\alpha$ and $\beta$. Fixing $\alpha=0.53$, we change $\beta$ to $0.254$ and, using the ``optimoptions" function in MATLAB, find the optimal regression parameter $\lambda=3\times 10^{-7}$. With the new parameters, the trained machine is able to predict the evolution of the Lorenz oscillator for about $9$ Lyapunov times (see Supplementary Material). The machine thus is regarded as properly trained. The size distribution of the synchronization clusters for this machine is plotted in Fig.~\ref{fig3}. We see that the distribution follows also the power-law scaling, with the fitted exponent identical to the one obtained for parameters $(\beta,\lambda)=(0.477,1\times 10^{-7})$. The similar results are also observed for the parameters $(\beta,\lambda)=(0.371,1\times 10^{-7})$ (by which the machine is also properly trained), as depicted in the same figure. We next check the impact of $\alpha$. For demonstration purpose, we choose $\alpha$ from different regimes in synchronization transition: $\alpha=0.25$ (from the desynchronization regime), $0.7$ (from the transition regime) and $0.8$ (from the strong synchronization regime). [See Fig.~\ref{fig1} for the locations of these parameters in synchronization transition and the size distributions of the synchronization clusters.] For each value of $\alpha$, we find the optimal parameters $(\beta,\lambda)$ by which the trained machine is able to predict the state evolution of the chaotic Lorenz oscillator for at least $6$ Lyapunov times. The other parameters of the RC are identical to the one used in Figs.~\ref{fig1}(a-c). The size distribution of the synchronization clusters for the three new machines are plotted in Fig.~\ref{fig3}. Interestingly, we see that all three distributions are ``shaped" to the same distribution: $p_s\propto s^{\gamma}$, with $\gamma\approx-1.7$. (See Supplementary Material for the fitting errors.) Again, it is found that when the machine is properly trained, the order parameter of the reservoir network is decreased. For instance, for the case of $\alpha=0.8$ shown in Fig.~\ref{fig3}, the order parameter is decreased from $0.952$ (the original network) to $0.038$ (the properly trained machine). The decreased order parameter is also attributed to the breaking of the giant clusters (Supplementary Material). More simulations have been conducted to check the generality of the above findings, including changing the size of the reservoir network, extending the time period for analyzing the size distribution of the clusters, and introducing noise perturbations to the phase oscillators. We find that, given the machine is properly trained, the same power-law scaling is always present in the reservoir. (See Supplementary Material for details.)

\begin{figure}[tbp]
\includegraphics[width=\linewidth]{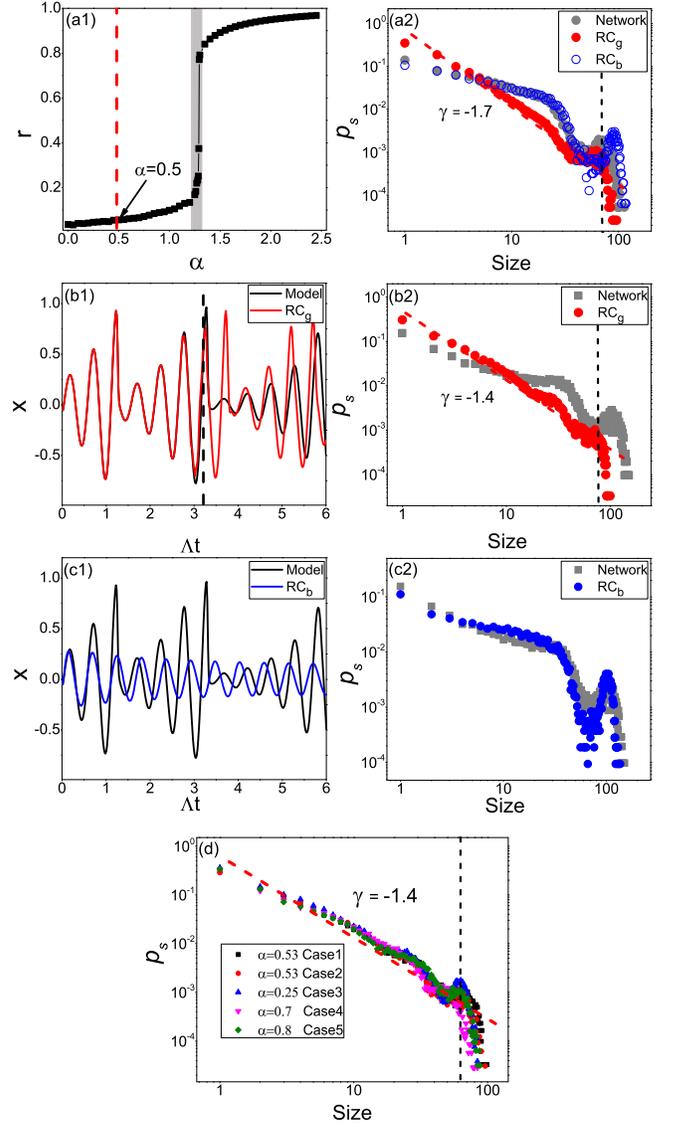}
\caption{The impacts of reservoir model, target dynamics and RC parameters on criticality. (a1) Synchronization transition for the conventional Kuramoto model. Grey zone denotes the transition regime. $r\approx 0.05$ at $\alpha=0.5$. (a2) By $\alpha=0.5$, the size distribution of the synchronization clusters for the original network without input (grey filled discs), the properly trained machine $RC_g$ (red filled discs) and the poorly trained machine $RC_b$ (blue open circles). The distribution for the properly trained machine follows the power-law scaling $p_s\propto s^{\gamma}$ with the fitted exponent $\gamma\approx -1.7$. (b1) For the chaotic R\"{o}ssler oscillator, the state evolution predicted by $RC_g$. The prediction horizon is about $3$ Lyapunov times. (b2) The size distribution of the synchronization clusters for $RC_g$. The distribution is fitted by a power-law scaling with exponent about $-1.4$. (c) The results for $RC_b$. In (b2) and (c2), the size distribution of the clusters for the original network (reservoir without input) is shown in grey. (d) For the chaotic R\"{o}ssler oscillator, the impacts of the RC parameters on criticality. Case 1: $\alpha=0.53$, $(\beta,\lambda)=(0.4,1\times 10^{-2})$ [the reference distribution copied from Fig.~\ref{fig2}(c)]. Case 2: $\alpha=0.53$, $(\beta,\lambda)=(0.366,1\times 10^{-2})$. Case 3: $\alpha=0.25$, $(\beta,\lambda)=(0.22,3\times 10^{-3})$. Cast 4: $\alpha=0.7$, $(\beta,\lambda)=(0.54,2\times 10^{-2})$. Case 5: $\alpha=0.8$, $(\beta,\lambda)=(0.512,1\times 10^{-2})$. Within the range $s\in [1,70]$, all distributions are fitted by the same power-law scaling $p_s\propto s^{\gamma}$, with $\gamma\approx -1.4$.}
\label{fig4}
\end{figure}

We finally check the impact of the reservoir model and the target dynamics on the properties of criticality. To check the impact of the network model, we replace the term $(1-\alpha)\omega_i$ with $\omega_i$ on the right-hand-side of Eq.~(\ref{reservoir-1}), so that the dynamics of the reservoir without the input is described by the conventional Kuramoto model~\cite{KuramotoModel}. The network size $N$, coupling matrix $\mathbf{A}$ and oscillator frequencies $\{\omega_i\}$ of the new model are identical to that of the generalized Kuramoto model. The variation of the order parameter $r$ with respect to the coupling strength $\alpha$ for the new model is plotted in Fig.~\ref{fig4}(a1). We see that the value of $r$ stays at small values till $\alpha\approx 1.2$, after which $r$ is increased quickly with $\alpha$ and reaches about $0.8$ at $\alpha=1.35$. The synchronization transition regime thus is approximately $\alpha\in (1.2, 1.35)$. Setting $\alpha=0.5$ (within the desynchronization regime) in the new model, we plot in Fig.~\ref{fig4}(a2) the size distribution of the synchronization clusters in the network (the grey discs), which is clearly different from the power-law scaling. Employing this network as the reservoir of RC, we train the RC by the time series of the chaotic Lorenz oscillator and use the trained RC to predict the future evolution. Still, we regard the machine as properly trained if it is able to predict the state evolution for $6$ Lyapunov times. Figure~\ref{fig4}(a2) plots the size distribution of the synchronization clusters for the properly ($RC_g$) and poorly ($RC_b$) trained machines. We see that the distribution of $RC_b$ is similar to that of the original network, but the distribution of $RC_g$ is ``shaped" to a power-law scaling $p_s\propto s^{\gamma}$, with the fitted exponent bing still about $-1.7$.  
 
We employ the chaotic R\"{o}ssler oscillator to check the impact of the target dynamics on the properties of criticality. The dynamics of the oscillator is governed by the equations: $(dx/dt,dy/dt,dz/dt)^{T}=[-y-z, x+0.2 y, 0.2+(x-9)z]^{T}$. The largest Lyapunov exponent of the oscillator is $\Lambda\approx 0.085$. Still, the state variables are normalized to be within the range $[-1,1]$. For the purpose of comparison, we set the reservoir to be identical to the one used in predicting Lorenz chaos in Fig.~\ref{fig2} (i.e., the generalized Kuramoto model with the parameters $N=500$ and $\alpha=0.53$). For the chaotic R\"{o}ssler oscillator, the machine is regarded as properly trained if it can predict the system evolution for $3$ Lyapunov times. An example of properly trained RC is shown in Fig.~\ref{fig4}(b1), in which the RC parameters are $(\beta,\lambda)=(0.4,0.01)$. The size distribution of the synchronization clusters for this machine is plotted in Fig.~\ref{fig4}(b2). We see that the distribution follows still a power-law scaling, but with a different exponent $\gamma\approx -1.4$. An example of poorly trained RC is shown in Fig.~\ref{fig4}(c1), in which the RC parameters are $(\beta,\lambda)=(0.027,0.01)$. We see that, comparing with the distribution of the original network (reservoir network without input), the distribution of the poorly trained machine is only slightly adjusted. Figure~\ref{fig4}(d) shows the size distribution of the synchronization clusters for different machines, all are properly trained and are able to predict the evolution of the R\"{o}ssler oscillator for at least $3$ Lyapunov times. We see that, despite the variation of $\alpha$, all the distributions fall onto the same power-law scaling $p_s\propto s^{\gamma}$ with $\gamma\approx -1.4$. These results are consistent with the results for Lorenz chaos (see Fig.~\ref{fig3}), where it is shown that the scaling exponent is also independent on the RC parameters.

The results in Figs.~\ref{fig3} and \ref{fig4} show the generality of criticality in machine learning, and also point out the impact of the target dynamics on criticality -- modifying the scaling exponent $\gamma$. The dependence of $\gamma$ on the dynamics of the target system is reasonable, as one of the necessary conditions for a proper training of RC is the establishment of generalized synchronization between the target system and the reservoir~\cite{RC:LZX2018,RC:LZX2020}. As such, the change of the dynamics of the target system must result in the change of the dynamics of the reservoir network, which, according to Figs.~\ref{fig3} and \ref{fig4}, seems to be reflected in the scaling exponent $\gamma$. 

Additional simulations have been conducted to check the impact of the target dynamics on the scaling properties. The results show that the scaling exponent is not affected by the bifurcation parameters of the target system. For instance, replacing the target system with a periodic Lorenz oscillator or another chaotic Lorenz oscillator (with different bifurcation parameters), the size distribution of the synchronization clusters in properly trained machines follows the same power-law scaling presented in Fig.~\ref{fig3}. Simulations have also been conducted for learning chaotic mapping system, in which a power-law scaling is also observed in the size distribution of the synchronization patterns, but with a different exponent. (See Supplementary Material for more details.) These findings are reminiscent of the transfer learning of chaotic systems~\cite{RC:FHW2021,KLW:2021,RC:Inubushi,RC:Guo2021}, where it is shown that the RC trained by the time series of a periodic oscillator can be used to infer the statistical properties of a chaotic oscillator with the same type of dynamics but different bifurcation parameters. The fact that knowledge can be transferred between different systems suggests that it is the intrinsic dynamics that the machine leans from the data, instead of the mathematical expressions describing the trajectories. Putting alternatively, from the standpoint of machine learning, the complexity of the training data may lie in the complexity of the dynamical functions of the target system, instead of the specific properties of the system dynamics such as dimension, Lyapunov exponents and bifurcation parameters. Our finding that the scaling properties of the machine is not affected by the bifurcation parameters of the target dynamics supports this understanding.  

Many questions remain open. One is about the relationship between criticality and prediction performance. Whereas our preliminary results show evidences of criticality in properly trained machines, the scaling relationships obtained are approximate and rough. To confirm the scaling relationships, sophisticated analysis should be conducted~\cite{Criti:Friedman2012,Criti:Fosque2021}. Also, it will be interesting to check whether the power-law distribution can be better fitted by improving the prediction performance. To study this, one may need to consider a large-size reservoir and tune the RC parameters very carefully. Another question concerns the reverse: if the reservoir is tuned artificially to the state of criticality characterized by a specific scaling exponent, will the prediction performance be optimized? As the scaling exponent is dependent on the type dynamics of the target system but not the bifurcation parameters, to generate a distribution satisfying a specific scaling exponent, one needs to scan the hyperparameter of the RC with a high precision. Both questions are worth pursuing, yet requiring extensive simulations. We hope these questions could be addressed by further studies.        

Summarizing up, a new reservoir of coupled phase oscillators has been designed and employed to explore the working mechanism of RC. It is found that despite of the synchronization status of the original reservoir, once the machine is properly trained, oscillators in the reservoir are synchronized in clusters, with the sizes of the clusters following a unique power-law distribution. The exponent of the distribution is independent on the system parameters, but is affected by the type of dynamics of the target system. 

This work was supported by the National Natural Science Foundation of China under the Grant Nos.~11875182, 11775034 and 11775035.

\end{document}